\documentclass[12pt,showpacs,preprintnumbers]{iopart}

\usepackage{amsmath,amssymb}
\usepackage{graphicx}
\usepackage{pstricks}
\usepackage{dcolumn}
\usepackage{bm}
\usepackage[latin1]{inputenc}

\newcommand{\bbr}{\bold{r}}
\newcommand{\bR}{\bold{R}}

\newcommand{\bU}{\bold{U}}
\newcommand{\R}{\mathbb{R}}
\newcommand{\N}{\mathbb{N}}
\newcommand{\Z}{\mathbb{Z}}

\newcommand{\cN}{\mathcal{N}}
\newcommand{\cF}{\mathcal{F}}
\newcommand{\cH}{\mathcal{H}}

\newcommand{\cE}{\mathcal{E}}
\newcommand{\ii}{\infty}
\newcommand\pscal[1]{{\ensuremath{\langle #1 \rangle}}}

\begin{document}


\title{Non-perturbative embedding of local defects in crystalline materials}

\author{Eric Canc\`es$^1$, Am\'elie Deleurence$^1$ and Mathieu Lewin$^2$}

\address{$^1$ CERMICS, \'Ecole des Ponts and INRIA, 6 \& 8 Av. Blaise
  Pascal, Cit\'e Descartes, 77455 Marne-la-Vall\'ee Cedex 2, France, \\
  cances@cermics.enpc.fr, deleurence@cermics.enpc.fr} 

\address{$^2$ CNRS \& Laboratoire de Mathématiques UMR 8088, Université de Cergy-Pontoise, 2 Avenue Adolphe Chauvin, 95302 Cergy-Pontoise Cedex, France,\\
  Mathieu.Lewin@math.cnrs.fr} 

\date{\today}

\begin{abstract}
We present a new variational model for computing the
electronic first-order density matrix of a crystalline material in
presence of a local defect. A natural way to obtain variational
discretizations of this model is to expand the difference $Q$ between the
density matrix of the defective crystal and the density matrix of the
perfect crystal, in a basis 
of precomputed maximally localized Wannier functions of the
  reference perfect crystal. This approach can be used within
any semi-empirical or Density Functional Theory framework. 
\end{abstract}

\pacs{
71.15.-m 
}
\maketitle


Describing the electronic state of crystals with local defects is a
major issue in solid-state physics, materials science and
nano-electronics~\cite{Pisani,Kittel,Stoneham}. The first self-consistent
electronic structure calculations for defective crystals were performed
in the late 70', by means of nonlinear Green functions
methods~\cite{GF1,GF2,GF3}.
In the 90', it became possible to solve the Kohn-Sham
equations~\cite{KohnSham} for systems 
with several hundreds of electrons, and Green function methods were
superseded by supercell methods~\cite{supercell,supercell2}. However,
supercell methods have several drawbacks. 
First, the defect interacts with its periodic images. Second, the
supercell must have a neutral total charge, so that in the simulation of
charged defects, an artificial charge distribution (a jellium for
instance) needs to be introduced to counterbalance the charge of the
defect. These two drawbacks may lead to large, uncontrolled
errors in the estimation of the energy of the defect. In practice, {\it
  ad hoc} correction terms are introduced to account for these
errors~\cite{MP}. A refinement of the supercell approach, based on a
more careful treatment of the Coulomb interaction, has also been 
proposed in~\cite{Schultz}.

In a recent article~\cite{CDL}, we have used rigorous thermodynamic limit
arguments to derive a variational model allowing to directly compute the
modification of the electronic first order density matrix generated by a
(neutral or charged) local defect, when the host crystal is an insulator
(or a semi-conductor). This model has a structure similar to the
Chaix-Iracane model in quantum electrodynamics~\cite{CI,HLS4}. 
This similarity originates from 
formal analogies between the Fermi sea of a defective crystal and the
Dirac sea in presence of atomic nuclei. For technical reasons, the
reference model considered in~\cite{CDL} was the reduced Hartree-Fock
model, or in other words, a Kohn-Sham model with fractional
occupancies and exchange-correlation energy set to zero.

The purpose of the present article is twofold. First, the extension of
our model to a generic exchange-correlation functional is
discussed. Second, a rigorous justification of the numerical method
consisting in expanding the difference between the density matrix of the
defective crystal and the density matrix of the perfect crystal, in a
basis of well-chosen 
Wannier functions of the reference perfect crystal, is provided: this
method can be seen as a variational approximation of our model.

\section{Derivation of the model}

We consider a generic Kohn-Sham model (or rather a generic {\em
  extended} Kohn-Sham model in which fractional occupancies are allowed)
with exchange correlation energy functional $E^{\rm xc}(\rho)$. 
For the sake of simplicity, we omit the spin variable. 
The ground
state of a molecular system with nuclear charge density~$\rho^{\rm nuc}$
and ${\mathcal N}$ electrons is obtained by solving
\begin{equation} \label{eq:KS_pb}
\inf \left\{ E^{\rm KS}_{\rho^{\rm nuc}}(\gamma), \; 0 \le \gamma \le 1,
  \; \Tr(\gamma) = {\mathcal N} \right\},
\end{equation}
\begin{equation} \label{eq:KS_energy}
E^{\rm KS}_{\rho^{\rm nuc}}(\gamma) = \Tr\left( -\frac 1 2 \Delta \gamma
\right) - D(\rho^{\rm nuc},\rho_\gamma) + \frac 1 2
D(\rho_\gamma,\rho_\gamma) + E^{\rm xc}(\rho_\gamma),
\end{equation}
where $\rho_\gamma(\bbr) = \gamma(\bbr,\bbr)$ and where
$$
D(f,g) = \int_{\R^3} \int_{\R^3} \frac{f(\bbr) \, g(\bbr')}{|\bbr-\bbr'|} \,
d\bbr \, d\bbr' 
$$
is the Coulomb interaction.
Still for simplicity, we detail the case of the X$\alpha$ exchange-correlation
functional 
$$
E^{\rm xc}(\rho) = -C_{{\rm X}\alpha} \int_{\R^3} \rho^{4/3}, 
$$
the extension to more accurate LDA functionals being
straightforward. Likewise, replacing the all electron model considered
here with a valence electron model with pseudopotentials does not bring
any additional difficulty.  

The above model describes a \emph{finite} system of $\cN$ electrons in
the electrostatic field created by the density $\rho^{\rm
  nuc}$. Our goal is to describe an \emph{infinite} crystalline material
obtained in the thermodynamic limit $\cN\to\ii$. In fact we shall
consider two such systems. The first one is the periodic crystal obtained when, in the thermodynamic limit, the nuclear density approaches the periodic nuclear distribution of the perfect
crystal:
\begin{equation}
 \rho^{\rm nuc}\rightarrow \rho_{\rm per}^{\rm nuc},
\label{case_periodic}
\end{equation}
$\rho^{\rm nuc}_{\rm per}$ being a periodic distribution. The second system
is the previous crystal in presence of a local defect: 
\begin{equation}
\rho^{\rm nuc}\rightarrow \rho_{\rm per}^{\rm nuc}+\nu. 
\label{case_defect}
\end{equation}
Typically, $\nu$ describes nuclear vacancies, interstitial nuclei, or
impurities together with possible local rearrangement of the nuclei of
the host crystal in the vicinity of the defect. In the simple case of a
reference perfect crystal with a single atom per unit cell
$$
\rho^{\rm nuc}_{\rm per} = \sum_{\bR \in {\cal R}}  z \delta_{\bR}
$$
where $\cal R$ is the Bravais lattice of the host crystal and
$\delta_{\bR}$ is the Dirac delta measure at $\bR$. If the defect
consists in a impurity (the nucleus of charge $z$ at $\bR={\bold 0}$
being replaced with a nucleus of charge $z'$), the charge distribution
$\nu$ reads 
$$
\nu = z' \delta_{\bU({\bold 0})} - z \delta_{\bold 0} + \sum_{\bR \in {\cal R}
  \setminus \left\{0\right\}} z \left( \delta_{\bR + \bU(\bR)} -
  \delta_{\bR} \right), 
$$ 
where $\bU$ is the displacement field of the nuclei generated by the
relaxation of the crystal. It is therefore composed of nuclei of
positive charges and of ``ghost nuclei'' of negative charges.
In this article, we assume that $\nu$ is
given, and we focus on the calculation of the electronic density 
matrix.

\begin{figure}[h]
\centering
\def\JPicScale{1.3} 
\input{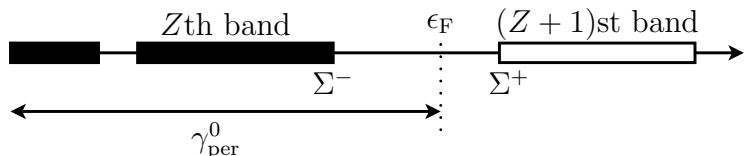}
\caption{Spectrum of $H^0_{\rm per}$.}
\label{fig:periodic}
\end{figure}

The form of the density matrix $\gamma^0_{\rm per}$ of the \emph{perfect crystal} obtained in the thermodynamic limit \eqref{case_periodic} is well-known.  
The matrix $\gamma^0_{\rm per}$ is a solution to the self-consistent equation
\begin{equation}
 \gamma^0_{\rm per}=\chi_{(-\ii;\epsilon_F]}(H^0_{\rm per})
\label{eq:SCF0}
\end{equation}
\begin{equation}
 H^0_{\rm per} = - \frac 1 2 \Delta + \Phi_{\rm per} - \frac 4 3 C_{{\rm X}\alpha}\,{\rho^0_{\rm
  per}} ^{\!\!\!\!\!1/3}, 
\label{def_H_0}
\end{equation}
$$
-\Delta \Phi_{\rm per} = 4 \pi \left( \rho_{\rm per}^0 - \rho_{\rm
    per}^{\rm nuc} \right), \quad \Phi_{\rm per}
\mbox{ ${\mathcal R}$-periodic}.
$$
The notation $P=\chi_{(-\ii;\epsilon_{\rm F}]}(A)$ means that $P$ is the
spectral orthogonal projector of the self-adjoint operator $A$ corresponding to
filling all the energies up to the Fermi level $\epsilon_{\rm F}$ (see
for instance~\cite{RS}). In our case, \eqref{eq:SCF0} means that $\gamma^0_{\rm per}$ is the spectral projector which fills all the energies of $H^0_{\rm per}$ up to the Fermi level $\epsilon_{\rm F}$, see Figure \ref{fig:periodic}.

The density of the periodic Fermi sea is $\rho^0_{\rm  per}(\bbr)=\gamma^0_{\rm per}(\bbr,\bbr)$. Note that the system is locally neutral:
$$\int_\Omega\rho^0_{\rm  per}=\int_\Omega\rho^{\rm nuc}_{\rm  per},$$
where $\Omega$ is a reference unit cell, 
the Fermi level $\epsilon_{\rm F}$ being chosen to ensure this
equality. For the rest of the article, we assume that the host crystal is an insulator (or a semi-conductor), i.e. that
there is a gap $g = \Sigma^+ - \Sigma^- > 0$ between the highest occupied and
the lowest virtual bands. Then the Fermi level can be any number $\Sigma^-\leq\epsilon_{\rm F}<\Sigma^+$.

Now we consider the system obtained in the thermodynamic limit \eqref{case_defect} when there is a defect $\nu$ and derive a nonlinear variational model for it.
We shall describe the variations of the Fermi sea with respect to the
periodic state $\gamma^0_{\rm per}$. The relevent variable therefore is
$$Q=\gamma-\gamma^0_{\rm per}$$
where $\gamma$ is the density matrix of the defective Fermi sea. 
Notice that the constraint that $\gamma$ is a density matrix ($0\leq
\gamma\leq1$) translates into $-\gamma^0_{\rm per} \le Q \le 
  1 - \gamma^0_{\rm per}$ for the new variable $Q$.

The energy of $Q$ is by definition the difference of two infinite
quantities: the energy of the state $\gamma$ and the energy of the
periodic Fermi sea $\gamma^0_{\rm per}$. Using \eqref{eq:KS_energy}, one
obtains:
\begin{equation} 
\cE^\nu(Q)  =  \Tr(H^0_{\rm per} Q) - D(\nu,\rho_Q)
 + \frac 1 2 
D(\rho_Q,\rho_Q) + \epsilon^{\rm xc}(\rho_Q) \label{eq:energy_formal}
\end{equation}
where 
$$
\epsilon^{\rm xc}(\rho_Q) = - C_{{\rm X}\alpha}  \int_{\R^3} 
(\rho^0_{\rm per} + \rho_Q)^{4/3} - {\rho^0_{\rm per}}^{\!\!\!\!\!4/3} -
\frac 4 3 {\rho^0_{\rm per}}^{\!\!\!\!\!1/3} \rho_Q. 
$$

If we want to describe a defective crystal of electronic charge $q$ ($q$
electrons in excess with respect to the perfect crystal if $q>0$, or $-q$ holes if $q < 0$) interacting with the self-consistent Fermi sea in the presence of the defect, we have to consider the minimization principle
\begin{equation} \label{eq:min_formal} 
E^\nu(q)=\inf \left\{ \cE^\nu (Q), \, -\gamma^0_{\rm per} \le Q \le
  1 - \gamma^0_{\rm per}, \; \Tr(Q) = q \right\}.
\end{equation}
We obtain in this way a model which apparently renders
possible the direct calculation of the defective Fermi sea
in presence of the nuclear charge defect $\nu$, when $q$ electrons (or
$-q$ holes) are trapped by the defect.
A globally neutral system would correspond to $q=\int_{\R^3}\nu$ but there is no obstacle in applying \eqref{eq:min_formal} to charged defects.

Alternatively, one can, instead of imposing {\it a priori} the total
charge $q$ of the system (microcanonical 
viewpoint), rather fix the Fermi level $\epsilon_{\rm F} \in
(\Sigma^-,\Sigma^+)$ (grand-canonical viewpoint). This amounts to
considering the Legendre transform of \eqref{eq:min_formal}: 
\begin{equation} \label{eq:min_legendre_formal} 
E^\nu_{\epsilon_{\rm F}}=\inf \left\{ \cE^\nu (Q)-\epsilon_{\rm F}\Tr(Q), \, -\gamma^0_{\rm per} \le Q \le
  1 - \gamma^0_{\rm per} \right\}.
\end{equation}

Any solution of \eqref{eq:min_formal} or \eqref{eq:min_legendre_formal} satisfies the SCF equation
\begin{equation} \label{eq:SCF}
Q = \chi_{(-\infty,\epsilon_{\rm F})}\left( H_Q\right)-\gamma^0_{\rm per} + \delta,
\end{equation}
where 
$$
H_Q = -\frac\Delta2 + \Phi_{\rm per} +  (\rho_Q-\nu) \ast
  \frac{1}{|x|} - \frac 4 3 C_{{\rm X}\alpha} (\rho^0_{\rm per} + \rho_Q)^{1/3}
$$
and where $0 \le \delta \le 1$ is a finite-rank self-adjoint operator on
$L^2(\R^3)$ such that $\mbox{Ran}(\delta) \subset
\mbox{Ker}(H_Q-\epsilon_{\rm F})$. In the case of \eqref{eq:min_formal},
the Fermi level $\epsilon_{\rm F}$ is the Lagrange multiplier associated
with the constraint $\Tr(Q)=q$.
The essential spectrum of $H_Q$ is the same as the one of $H^0_{\rm per}$
and is therefore composed of bands. On the other hand, the discrete spectrum
of $H^0_{\rm per}$ is empty, while the discrete spectrum of $H_Q$ may
contain isolated eigenvalues of finite multiplicities located below the
essential spectrum and between the bands. Each filled (or unfilled) eigenvalue may correspond to electrons (or holes) which are trapped by the defect.

\begin{figure}[h]
\centering
\def\JPicScale{1.3} 
\input{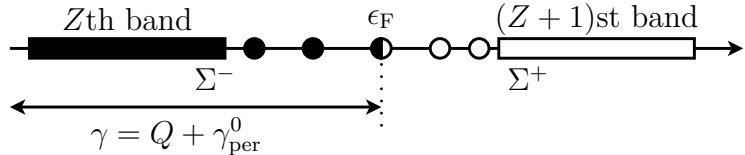}
\caption{Spectrum of $H_Q$.}
\label{fig:defaut}
\end{figure}

The SCF equation \eqref{eq:SCF} is equivalent to the usual Dyson
equation, which is at the basis of Green function methods.

\section{Proper definition of the variational set}

The variational models~\eqref{eq:min_formal} and
\eqref{eq:min_legendre_formal} may look similar to the usual Kohn-Sham
models for molecules and perfect crystals. Their mathematical structure
is however dramatically more complex. To design consistent
numerical methods for solving~\eqref{eq:min_formal} and
\eqref{eq:min_legendre_formal}, a deeper understanding of the
mathematical setting is needed.

The biggest issue with problems \eqref{eq:min_formal} and
\eqref{eq:min_legendre_formal} is to properly define the variational
set, that is the set of all $Q$'s on which one has to minimize the
energy functional ${\mathcal E}^\nu(Q)$ or the free energy functional 
${\mathcal E}^\nu(Q) - \epsilon_{\rm F} \Tr(Q)$. For usual Kohn-Sham
models, the variational set is very simple: it is the largest set of
density matrices for which each term of the energy functional is a
well-defined number and the constraints are satisfied. This
is the reason why it is not a problem to omit the precise definition of
the variational set when dealing with usual Kohn-Sham models. For
instance, the variational set for \eqref{eq:KS_pb} is
\begin{equation} \label{eq:def_var_KS_N}
\left\{ \gamma \; | \; 0 \le \gamma \le 1, \; \Tr(\gamma) = {\mathcal N}, \; 
\Tr( |\nabla| \gamma |\nabla|) < \infty \right\}.
\end{equation}
Let us recall (see \cite{RS} for instance) that if $B$ is a non-negative
self-adjoint operator on $L^2(\R^3)$ and if $(\psi_i)_{i \in
  \N}$ is an orthonormal basis of $L^2(\R^3)$, the series of non-negative
numbers $\sum_{i=0}^{+\infty} \langle \psi_i |B| \psi_i
\rangle$ converges in $\R_+ \cup \left\{+\infty \right\}$ towards a limit
  denoted by $\Tr(B)$, which does not depend on the chosen basis. The
  operator $B$ is said to be trace-class if $\Tr(B) < \infty$.
A bounded operator $A$ on $L^2(\R^3)$
is trace-class if $\sqrt{A^\ast A}$ is trace-class. In this
case, the scalar
$\Tr(A) = \sum_{i=0}^{+\infty} \langle \psi_i |A| \psi_i \rangle$ is
well-defined and does not depend on the chosen basis. 
On the other hand, if $A$ is not trace-class, the series
$\sum_{i=0}^{+\infty} \langle \psi_i |A| \psi_i \rangle$ may converge
for one specific basis and diverge (or converge to a different limit) in
another basis.

The condition $\Tr( |\nabla| \gamma |\nabla|) < \infty$
in~\eqref{eq:def_var_KS_N} is a 
necessary and sufficient condition for each term of \eqref{eq:KS_energy}
being well-defined. In terms of Kohn-Sham orbitals, this conditions
means that each 
orbital $\phi_i$ is in the Sobolev space $H^1(\R^3) = \left\{ \phi \in
  L^2(\R^3) \; | \; \nabla \phi \in (L^2(\R^3))^3 \right\}$.

The difficulty with the variational models~\eqref{eq:min_formal} and
\eqref{eq:min_legendre_formal} is that the variational set has not
so simple a structure. It
was shown in \cite{CDL} that an appropriate variational set is
the convex set 
\begin{multline*}
{\mathcal K}  =  \big\{ Q\ |  \ -\gamma^0_{\rm per} \le Q \le 1 - 
\gamma^0_{\rm per},\ \Tr(1+|\nabla|)Q^2(1+|\nabla|)\\
 +\Tr(1+|\nabla|)(Q^{++}-Q^{--})(1+|\nabla|)<\ii \big\}.
\end{multline*}
In the above expression, we have used the notation 
$$
Q = \quad \left( 
\begin{array}{c||c}
Q^{--} & Q^{-+}  \\
\hline \hline 
Q^{+-} & Q^{++}  \\
\end{array}
\right) 
$$
with
$$
Q^{--} = \gamma^0_{\rm per} Q \gamma^0_{\rm per}, \; 
 \; 
Q^{-+} = \gamma^0_{\rm per} Q (1-\gamma^0_{\rm per}),
$$
$$
Q^{+-} = (1-\gamma^0_{\rm per}) Q \gamma^0_{\rm per} , \; 
 \; 
Q^{++} = (1-\gamma^0_{\rm per}) Q (1-\gamma^0_{\rm per}),
$$
corresponding to the decomposition 
\begin{equation} \label{eq:decomp}
L^2(\R^3) = {\cal H}_-\oplus{\cal H}_+,
\end{equation}
where ${\mathcal H}_-=\gamma^0_{\rm per}L^2(\R^3)$ and ${\mathcal H}_+
=(1-\gamma^0_{\rm per})L^2(\R^3)$ are respectively the occupied and
virtual spaces of the reference perfect crystal. 

Notice that when $Q$ satisfies the
constraint $-\gamma^0_{\rm per} \le Q \le 1 - 
\gamma^0_{\rm per}$, one has $Q^{++} \ge 0$ and  $Q^{--} \le 0$. 
A remarkable point, proved in~\cite{CDL}, is that the density $\rho_Q$
of any operator $Q\in\mathcal{K}$ is a well-defined function which
satisfies  
$$
\int_{\R^3}\rho_Q^2+D(\rho_Q,\rho_Q)<\ii.
$$
This shows that the electrostatic components of the energy
${\mathcal E}^\nu(\gamma)$ are well-defined and that so is the
exchange-correlation contribution:
as $\rho^0_{\rm per}$ is periodic, continuous and  
positive on $\R^3$ and as $\rho_Q \in L^2(\R^3)$, the fifth term of
\eqref{eq:energy_formal} which was not considered in \cite{CDL} is also
well-defined. Finally, following \cite{HLS1}, the generalized trace of
an operator $Q\in\mathcal{K}$ is defined by 
\begin{equation}
\Tr(Q)=\Tr(Q^{++})+\Tr(Q^{--}),
\label{def_charge}
\end{equation}
and for any $Q \in \mathcal{K}$, one sets
$$
\Tr(H^0_{\rm per}Q)=\Tr([H^0_{\rm per}]^{++}Q^{++})
+\Tr([H^0_{\rm per}]^{--}Q^{--}),
$$
where $[H^0_{\rm per}]^{--}$ and $[H^0_{\rm per}]^{++}$ are respectively
the restrictions  to the occupied and virtual spaces of the periodic
Kohn-Sham hamiltonian of the perfect crystal. Note that $H^0_{\rm per}$ is
block diagonal in the decomposition~\eqref{eq:decomp}:
$$
H^0_{\rm per} = \quad \left( 
\begin{array}{c||c}
[H^0_{\rm per}]^{--} & 0  \\
\hline \hline 
0 & [H^0_{\rm per}]^{++}  \\
\end{array}
\right) .
$$
The definition~\eqref{def_charge} of the trace function is an extension
of the standard 
trace function defined on the set of trace-class operators. Note that
this extension depends of $\gamma^0_{\rm per}$
through the decomposition~\eqref{eq:decomp} of the $L^2$ space.
In the Quantum Electrodynamical model studied in
\cite{HLS1,HLS2,HLS3,HLS4,HLSo}, minimizers are never
trace-class (this property being related to renormalization). Whether or
not the minimizers of ~\eqref{eq:min_formal} and
\eqref{eq:min_legendre_formal} are trace-class still is an open question.

To our knowledge, the variational interpretation of the ground state
solutions of the self-consistent equation \eqref{eq:SCF} as minimizers
of the energy 
\eqref{eq:energy_formal} on the set ${\cal K}$ with a constraint on the
generalized trace \eqref{def_charge}, is new. 
This interpretation allows to rigorously justify the numerical method
described in Section~\ref{sec:var_app}. 

\section{Interpretation in terms of Bogoliubov states}

The density matrix formalism used in the previous section can be
reinterpreted in terms of Bogoliubov states, following~\cite{CI}.

Let $\gamma$ be an
orthogonal projector acting on $L^2(\R^3)$ such that
$Q = \gamma-\gamma^0_{\rm per} \in {\mathcal K}$. It can be proved
\cite{HLS3} that there exists an orthonormal basis
$(\phi_i^-)_{i\geq-N_-}$ of ${\cal H}_-$ 
and an orthonormal basis $(\phi_i^+)_{i\geq-N_+}$ of ${\cal H}_+$
  such that in this basis
\begin{equation}
 Q = \quad \left( 
\begin{array}{c|c||c|c}
- I_{N_-} & 0 & 0 & 0 \\
\hline 
0 & \mbox{diag}(-p_i) & 0 & \mbox{diag}(p_i') \\
\hline \hline 
0 & 0 & I_{N_+} & 0 \\
\hline 
0 & \mbox{diag}(p_i') & 0 & \mbox{diag}(p_i) \\
\end{array}
\right) 
\label{decomp_Q}
\end{equation}
with $0 \le p_i < 1$, $\sum_{i=0}^{+\infty} p_i < \infty$, $p_i' =
\sqrt{p_i(1-p_i)}$. Notice that $Q$ is a
trace-class operator if and only if $\sum_{i=0}^{+\infty} \sqrt{p_i} < \infty$.
Let us assume for simplicity that in equation~\eqref{eq:SCF}, the Fermi
level $\epsilon_{\rm F}$ is either empty or fully occupied. In this
case, $\chi_{(-\infty,\epsilon_{\rm F})}\left( H_Q\right) + \delta$ is
an orthogonal projector, which implies that $Q$ can be decomposed as
in~\eqref{decomp_Q}. It is important to mention that in this case, the
generalized trace of $Q$ is the integer $N_+ -N_-$.

Formula \eqref{decomp_Q} can be interpreted in terms of Bogoliubov
states. The orbitals $\phi_{-N_+}^+, \cdots, \phi_{-1}^+$ describe bound
electrons in the virtual bands of the reference perfect crystal, while the orbitals $\phi_{-N_-}^-, \cdots,\phi_{-1}^-$ represent bound holes in the occupied
bands. Likewise, each pair $(\phi_i^+,\phi_i^-)$ with $i\geq0$ and $0 <
p_i < 1$ is a virtual electron-hole pair, and $\phi_i^+$ and $\phi_i^-$
are the states of the corresponding Bogoliubov quasiparticles. The angle
$\theta_i=\mbox{asin}(p_i)$ is then called the Bogoliubov angle of the
virtual pair.

Formula \eqref{decomp_Q} can itself be rewritten in a second quantized
form, using the Fock space built upon the decomposition
(\ref{eq:decomp}). Let us introduce the $N$-electron sector
$\cF_+^N:=\bigwedge_1^N\cH_+$ and the $M$-hole sector
$\cF_-^M:=\bigwedge_1^M\cH_-$. The electron-hole Fock space is defined
as 
$$
\cF:=\bigoplus_{N,M\geq0}\cF_+^N\otimes\cF_-^M.
$$ 
We denote by
$a_i^\dagger$ the creation operator of an electron in the state
$\phi^+_i$ and by $b_i^\dagger$ the creation operator of a hole in the
state $\phi^-_i$. In this formalism, the vacuum state
$\Omega_0=1\otimes1\in\cF_+^0\otimes\cF^0_-$ corresponds to the periodic
Fermi sea of the perfect crystal, represented by the density matrix
$\gamma^0_{\rm per}$ in the usual Kohn-Sham description. 
We may also define the charge operator acting on the Fock space $\cF$ by
$${\cal Q}=\sum_{i\geq-N_+} a_i^\dagger a_i - \sum_{i\geq-N_-} b_i^\dagger b_i.$$

There is a special subclass of states in $\cF$ called Bogoliubov states
\cite{CI,HLS1,Berezin,BLS}. Each Bogoliubov state $\Omega_\gamma\in\cF$
is completely characterized by its one-body density matrix $\gamma$, an
orthogonal projector acting on $L^2(\R^3)$. Conversely, any projector
$\gamma$ gives rise to a Bogoliubov state under the Shale-Stinespring
\cite{Rui,ShSt} condition that $Q=\gamma-\gamma^0_{\rm per}$ is a
Hilbert-Schmidt operator (which means $\Tr(Q^2) < \infty$). The role of
the Shale-Stinespring condition is 
to ensure that $\Omega_\gamma$ is a well-defined state in the same Fock
space as the vacuum state $\Omega_0$. Saying differently, this ensures
that the Fock space representation associated with the splitting
$L^2(\R^3)=\gamma L^2(\R^3)\oplus (1-\gamma)L^2(\R^3)$ is equivalent to
the one induced by \eqref{eq:decomp} (i.e. $L^2(\R^3) = \gamma^0_{\rm per}
L^2(\R^3)\oplus (1-\gamma^0_{\rm per})L^2(\R^3)$). Notice the Hilbert-Schmidt 
condition $\Tr(Q^2)<\ii$ is satisfied for any $Q=\gamma-\gamma^0_{\rm
  per}$ in $\mathcal{K}$. Hence the variational set $\cal K$ can be identified with a variational set of Bogoliubov states $\{\Omega_\gamma\}_{\gamma\in\cal K}$ in the Fock space $\cal F$.

The expression of the Bogoliubov state $\Omega_\gamma$ in the Fock space
$\cF$ is given by~\cite{BLS,Rui,SS}
$$\Omega_\gamma=c\; a^\dagger_{-N_+}\cdots a^\dagger_{-1}b^\dagger_{-N_-}\cdots b^\dagger_{-1}\exp\left(\sum_{i\geq0}\lambda_i a^\dagger_ib^\dagger_i\right)\Omega_0$$
where $\lambda_i=\tan(\theta_i)$, and where $c$ is a normalization
constant. The above expression can be considered as the
second-quantized formulation of \eqref{decomp_Q}.
It can then easily be checked \cite{HLS1} that the charge of each
Bogoliubov state $\Omega_\gamma$ (counted relatively to that of the
vacuum $\Omega_0$) is actually given by \eqref{def_charge}:
$$\pscal{\Omega_\gamma|{\cal Q}|\Omega_\gamma}=\Tr(Q^{++})+\Tr(Q^{--})=N_+-N_-$$
where $Q=\gamma-\gamma^0_{\rm per}$.

\section{Variational approximation}
\label{sec:var_app}

Let us now come to the discretization of problem~\eqref{eq:min_formal}.

If one discretizes \eqref{eq:min_formal} in a local basis without taking
care of the constraint $Q\in{\cal K}$, there is a risk to
obtain meaningless numerical results. On the other hand, selecting a basis set
which respects the decomposition~\eqref{eq:decomp},
will lead to a well-behaved variational approximation of
\eqref{eq:min_formal} (the constraint $Q\in{\cal K}$ will be implicitly
taken into acount).
Let $V_\pm^h$ be finite-dimensional subspaces of the occupied and
virtual spaces ${\cal H}_\pm$ of the reference perfect crystal. Consider the
finite-dimensional subspace $V^h = V^h_- \oplus V^h_+$ of $L^2(\R^3)$,
the latter decomposition being the finite-dimensional counterpart of 
(\ref{eq:decomp}). Let
$(\phi_1, \cdots , \phi_{m_-})$ (resp. $(\phi_{m_-+1}, \cdots ,
\phi_{N_b})$) be an orthonormal basis of $V_-^h$ (resp. of
$V_+^h$). We denote for simplicity $m_+:=N_b-m_-$. The approximation set for $Q$ consists of the finite-rank operators
\begin{equation} \label{eq:approx_Q}
Q = \sum_{i,j=1}^{N_b} Q_{ij}^h |\phi_i \rangle \langle \phi_j|
\end{equation}
with $Q^h \in
\mathcal{K}^h  =  \big\{ Q^h=[Q^h]^T, \ 0 \le {\mathcal I} + Q^h \le 1
\big\}$, 
where ${\mathcal I}$ is the $N_b \times N_b$ block diagonal matrix 
$$
{\mathcal I} = \left[ \begin{array}{cc} 1_{m_-} & 0 \\ 0 & 0_{m_+}
  \end{array} \right] .
$$
The matrix of $H^0_{\rm per}$ in the basis $(\phi_i)$ is of the
form
$$
H^h = \left[ \begin{array}{cc} H^{--} & 0 \\ 0 & H^{++}
  \end{array} \right].
$$
For $Q$ of the form \eqref{eq:approx_Q}, it holds
$$
\cE^\nu(Q)  = \cE^{\nu}_h (Q^h)
$$
with 
$$
\rho_{Q^h}(r) = \sum_{i,j=1}^{N_b} Q^h_{ij} \phi_i(r) \, \phi_j(r)
$$
and
$$
\cE^{\nu}_h (Q^h)  =  \Tr(H^h Q^h) - D(\nu,\rho_{Q^h}) 
+ \frac 12 D(\rho_{Q^h},\rho_{Q^h}) + \epsilon^{\rm xc}(\rho_{Q^h}). 
$$
We then end up with the finite-dimensional optimization problem
\begin{equation} \label{eq:disc_VP}
E^{\nu}_h(q)=\inf \left\{ \cE^{\nu}_h (Q^h), \; Q^h \in {\mathcal K}^h,
  \; \Tr(Q^h) = q  \right\}
\end{equation}
which is a variational approximation of
\eqref{eq:min_formal}: 
$$E^{\nu}_h(q) \geq E^\nu(q).
$$

As $Q^h \in {\mathcal K}^h$ with $\Tr(Q^h) = q$ if and only if 
$${\mathcal I}+Q^h \in
 \left\{ D=D^T \in \R^{2N_b}, \, D^2 \le D, \, \Tr(D) = q+N_- \right\},$$
 problem
\eqref{eq:disc_VP} can be solved using relaxed constrained
algorithms~\cite{SCF1,SCF2}.

 The question is now to build spaces $V_-^h$ and
$V_+^h$ that provide good approximations
to~\eqref{eq:min_formal} and~\eqref{eq:min_legendre_formal}. A natural
choice is to use the  
maximally localized (generalized) Wannier functions~\cite{MV} (MLWFs) of
the reference perfect 
crystal. A very interesting feature of these basis functions is that they can
be precalculated once and for all for a given
  host crystal, independently of the local defect under consideration.
To construct $V_-^h$,
one can select the maximally localized (generalized)
Wannier functions 
of the occupied bands, that overlap with e.g. some ball $B_{R_c}$ of radius
$R_c$ centered on the nuclear charge defect. Note that due to the
variational nature of the approximation scheme, enlarging the
radius $R_c$ systematically improves the quality of the approximation.
To obtain a basis set 
for $V_+^h$, one can select a number of active (unoccupied) bands using an
energy cut-off and retain the maximally localized (generalized) Wannier
functions of the active bands that overlap with the same ball $B_{R_c}$.
The so-obtained basis set of the virtual space can be enriched by
adding projected atomic orbitals of the atoms and ghost atoms involved
in $\nu$ (using the localized Wannier functions of the occupied bands to
project out the ${\cal H}_-$ component of atomic orbitals preserves the 
locality of these orbitals).

\section{Numerical results}

In order to illustrate the efficiency of the variational approximation
presented above, we take the example of a one-dimensional (1D) model
with Yukawa interaction potential, for which the energy functional reads   
$$
E_{\rm 1D}(\gamma) = \Tr \left( - \frac 12 \, \frac{d^2\gamma}{dx^2} 
\right) - D_\kappa(\rho_{\rm nuc},\rho_\gamma) + \frac 1 2
D_\kappa(\rho_\gamma, \rho_\gamma)
$$
with
$$D_\kappa(f,g) = (A/2\kappa) \, \int_\R \int_\R f(x) \, {e^{-\kappa \, |x-x'|}}
\, g(x') \, dx \, dx'.$$
In the numerical examples reported below, the host crystal is
$\Z$-periodic and the nuclear density is a Dirac comb, i.e. 
$$\rho_{\rm nuc} = Z \sum_{j \in \Z} \delta_j,$$
with $Z$ a positive integer. The values of the parameters ($A=10$ and $\kappa=5$)
have been chosen in such a way that the ground state kinetic and
potential energies are of the same order of magnitude. 

\begin{figure}[h]
\centering
\includegraphics[width=10cm,angle=0]{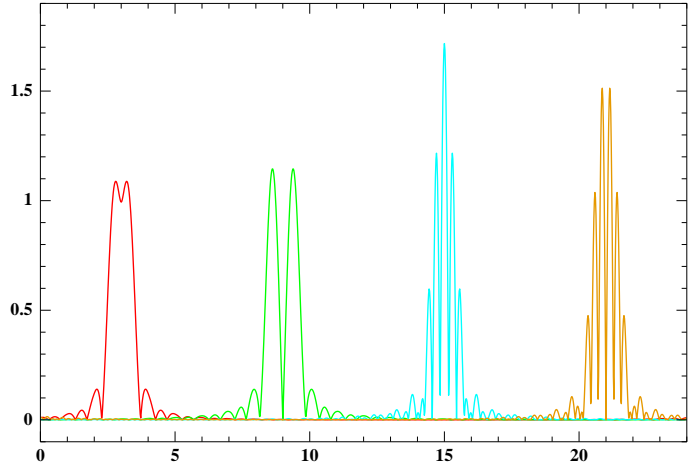}
\caption{Modulus of MLWFs
  associated with the two occupied bands (left) and with the lowest two
  virtual bands (right).}
\label{fig:MLWF}
\end{figure}
\begin{figure}[h]
\centering
\includegraphics[width=10cm,angle=0]{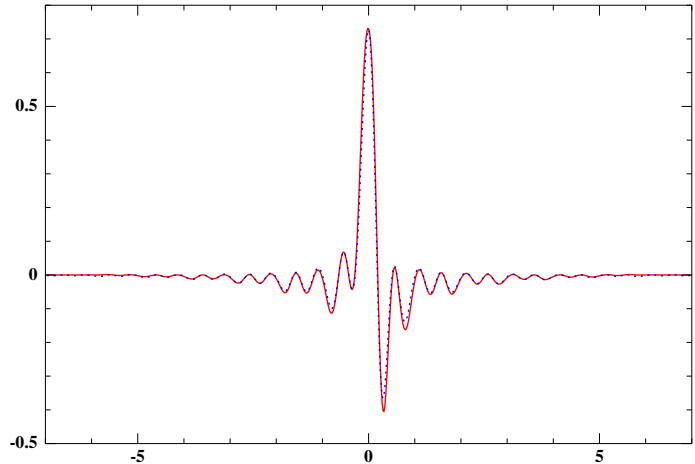}
\caption{Density $\rho_{Q^h}$ obtained with 28 MLWFs (line in red). The
  reference is a supercell calculation in a basis set of size~1224
  (dashed line in blue).} 
\label{fig:defect}
\end{figure}

The nuclear local
defect is taken of the form 
$$\nu = (Z-1)\delta_{0.25}-Z\delta_0.$$
This corresponds to moving one nucleus and lowering its charge by one unit. 

The first stage of the calculation consists in solving the cell
problem. For simplicity, we use a uniform discretization of the
Brillouin zone $(-\pi,\pi]$, and a plane wave
expansion of the crystalline orbitals.

The second stage is the construction of MLWFs. For this purpose, we make
use of an 
argument specific to the one-dimensional case~\cite{Resta}: the MLWFs
associated with the spectral projector $\gamma$ are the eigenfunctions
of the operator $\gamma x \gamma$. One first constructs $N_e$ mother MLWFs
(taking $\gamma = \gamma^0_{\rm per}$), then $N_a$ mother MLWFs
corresponding to the lowest $N_a$ virtual bands (taking for $\gamma$ the
spectral projector associated with the lowest $N_a$ virtual bands). The
so-obtained mother MLWFs are represented on Fig.~\ref{fig:MLWF}.

The third stage consists in constructing a basis set $(\phi_{j})_{1 \le j
  \le N_b}$ of $N_b=N_v(N_e+N_a)$ MLWFs by selecting the $N_v$
translations of the $(N_e+N_a)$ mother MLWFs that are closest to the
local defect, and in computing the first-order density matrix of the
form (\ref{eq:approx_Q}) which satisfies the constraints and minimizes
the energy. The profile of the density $\rho_{Q^h}$ obtained with 
$Z =2$, $N_e = 2$, $N_a=2$ and $N_b=28$ is displayed on
Fig.~\ref{fig:defect}. It is compared with a reference 
supercell calculation with 1224 plane wave basis functions. A fairly good
agreement is obtained with very few MLWFs. 

The implementation of our method in the Quantum Espresso suite of
programs~\cite{QuantumEspresso}, in the true 3D Kohn-Sham setting, is
work in progress~\cite{DCL}.   


\ack

This work was partially supported by the ANR grants LN3M and
  ACCQUAREL. A.D. acknowledges financial support from R\'egion
  Ile-De-France.


\end{document}